\pdfoutput=1
\documentclass[aps,pre,reprint,superscriptaddress,showpacs,preprintnumbers,amsmath,amssymb,showkeys]{revtex4-1}
\usepackage{amssymb}
\usepackage{amsmath}
\usepackage{graphicx}
\usepackage{color}
\usepackage{soul}
\usepackage{float}

\begin{document}

\title{Local variation of hashtag spike trains and popularity in Twitter}




\author{Ceyda Sanl{\i}}
\email{cedaysan@gmail.com(permanent); ceyda.sanli@unamur.be}
\affiliation{CompleXity and Networks, naXys, Department of Mathematics, University of Namur, 5000 Namur, Belgium}

\author{Renaud Lambiotte}
\email{renaud.lambiotte@unamur.be}
\affiliation{CompleXity and Networks, naXys, Department of Mathematics, University of Namur, 5000 Namur, Belgium}

\date{\today}

\begin{abstract}
We draw a parallel between hashtag time series and neuron spike trains. In each case, the process presents complex dynamic patterns including temporal correlations, burstiness, and all other types of nonstationarity.  We propose the adoption of the so-called local variation in order to uncover salient dynamics, while properly detrending for the time-dependent features of a signal. The methodology is tested on both real and randomized hashtag spike trains, and identifies that popular hashtags present regular and so less bursty behavior, suggesting its potential use for predicting online popularity in social media. 
\end{abstract}


\maketitle

\section*{Introduction}

The statistical properties of Twitter and, more generally, of human activity, are characterized by a strong heterogeneity in different dimensions. First, human behavior is known to generate bursty temporal patterns, significantly deviating from independent Poisson processes, as a majority of events take place over short time scales while a few events take place over very large times. This property translates into fat-tailed distributions for the timings $\Delta\tau$ between occurrences of a certain type of events, e.g. between two phone calls or two emails emitted by an individual. For instance, the inter-event time distribution $P(\Delta\tau)$ for the timings between two tweets of a user, or the use of a hashtag is well fitted by a power law such as $P(\Delta\tau)\approx\Delta\tau^\alpha$~\cite{scientificrumor}. The deviation from an exponential (uncorrelated) distribution may be either driven by complex decision-making and cascading mechanisms~\cite{BarbarasiOriginofBursts, Competition_memes, cooperationandcompetition} or by the time dependency of the underlying process, partly because of its intrinsic circadian and weekly rhythms~\cite{Amaral, plane}, as described in Fig. 1, or by a combination of these factors~\cite{Circadianpatternburstiness_mobilephone, burstydynamicsTwitter, 2014arXiv1411.0722F, 2015arXiv150203224M}. Importantly, the nonstationarity of the signal is known to broaden $P(\Delta\tau)$ and therefore to artificially increase the value of standard metrics, such the variance or the Fano factor, originally defined for stationary processes. Recently, a stochastic model for a stationary process also suggests a broad distribution in online user activity level on long time scales, longer than $\Delta\tau$~\cite{2015arXiv150203224M}.

\begin{figure}[]
\includegraphics[width= 10cm]{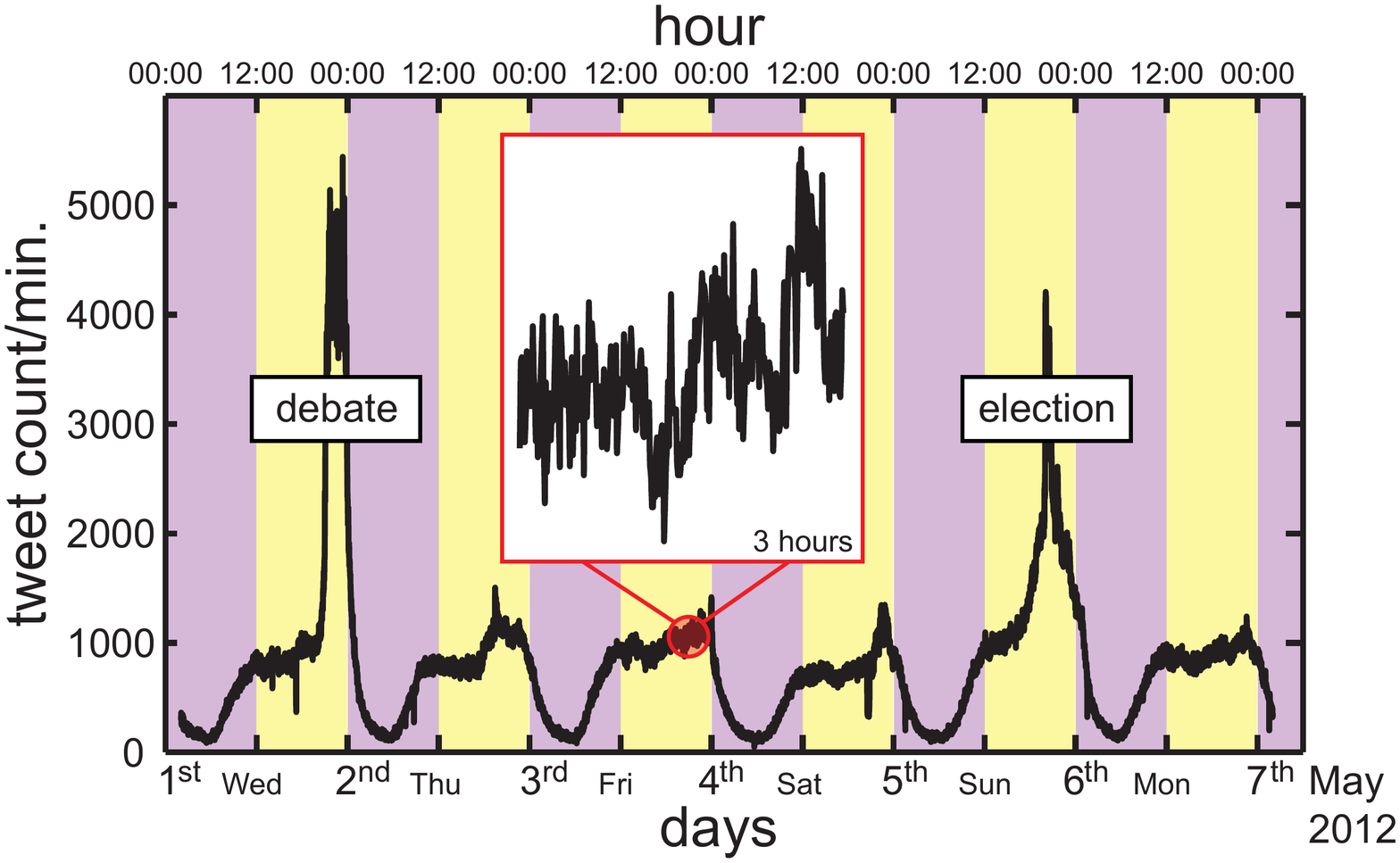}
\caption{Circadian pattern of tweeting activity. Increasing amount of tweets from midday (12:00) to midnight (00:00) is shown in the yellow shaded regions. Significant decays of activity are observed during nights. Activity increases during mornings as shown in purple shaded rectangles. In the inset, we show the temporal evolution at a finer scale, where fluctuations are visible. The data exhibits two peaks of activity in the evening of a political debate, on May 2 2012 (first peak) and on election day, May 6 2012 (second peak).}
\end{figure}

In addition to temporal heterogeneity in $\Delta\tau$, online human activity often generates a heterogeneity in popularity~\cite{onlinepopularityheterogeneity}. In the following, we focus on the popularity of hashtags in Twitter. Twitter is a micro-blogging service allowing users to post short messages, and to follow those published by other users. Messages often incorporate hashtags, keywords identified by the symbol $\#$, which users can track and respond to the message content and makes the platform interactive. Hashtags play a significant role in information diffusion by enhancing information and rumor spreading and consequently increase the impact of news. Discussions on protests~\cite{spanishprotestPLoSONE, spanishprotestSciRep} and political elections, advertisement of new products in marketing, announcements of scientific innovations~\cite{scientificrumor}, panic events such as earthquakes~\cite{earthquakeTwittercollectiveattention}, and comments on TV shows are some examples where hashtags are widely used. Additionally, hashtags can be even used to track and locate crisis~\cite{determinecrisisviahash} and can spread under the influences of both endogeneous factors, that is the propagation between Twitter users following each others, and exogeneous sources such as TV and newspapers~\cite{Dynamicsofbooksale}.

The popularity $p$ of a hashtag is measured by the number of times that it appears in an observation time window. While a majority of hashtags attracts no attention only very few of them propagate heavily~\cite{Viralityprediction_memes, Competition_memes}. Understanding the mechanisms by which certain hashtags or messages gain attention is a central topic of research in the study of online social media~\cite{leskovec}. Potential mechanisms for the emergence of this heterogeneity include forms of preferential attachment and competition-induced forces~\cite{Competition_memes_limitedattention, Competition_inducedcriticality, Competition_advertisement, 2015arXiv150105956G} driven by the limited amount of attention of users. 

Our main purpose is to explore connections between temporal and popularity heterogeneity. As a first contribution, we introduce a temporal measure for online human dynamics, suited for the analysis of nonstationary time series to quantify bursts, regularity, and temporal correlations. Originally defined for the study of inter-spike intervals of neurons~\cite{Lv1, Lv2, Lv3, Lv4, Lv5}, the so-called local variation $L_V$ is shown to identify and characterize deviations from Poisson (uncorrelated) processes, and to help predict successful hashtags.

\section*{Data mining and basic analysis}
\subsection*{Data collection and basic overview}

The data set has been collected via the publicly open Twitter streaming API between April 30, 2012, 10 pm and May 10, 2012, 10 pm. Only the geographical constraint has been applied as follows: The actions of all Twitter users located in France have been considered in order to avoid the existence of time differences between countries and regions, and no language filtering has been applied. The time resolution is 1 second and multiple activity can be recorded in the same second. During this time period, two major public events took place: An important political debate held on May 2 and the French presidential election-2012 held on  May 6. These events are not the topic of this work, but they are clearly visible in the time series, as shown in Fig. 1. 

The total number of tweets, including retweets, captured during the data collection is 9,747,351. The total number of tweets including at least one hashtag is 2,942,239. Around $30\%$ of the tweets therefore contain a hashtag. The fact that hashtags are used in regular tweets or in retweets is not specified. Moreover, any message (identical or not) considering at least one hashtag is recorded. Due to the debate and the election taking place during the data collection, the most popular hashtags are related to politics, as seen in Table 1. The time series of the hashtag study in this paper are provided in Supporting
Information S1. A total number of 473,243 individual users has been identified. Among those, 228,525 users published at least one hashtag, e.g. almost half of the social network is associated with hashtag diffusion.
In order to further characterize the importance of hashtags in Twitter activity, we compare the total number of seconds when any action is performed in the data set, 763,262 s $\approx$ 8.8 days and thus  $88\%$ of the total duration, to the number of seconds when at least one hashtag is published, 667,996 s $\approx$ 7.7 days, that is $ 77\%$ of the total duration. In any case, the hashtag data cover a majority of the time window, even during off-peak hours. These numbers confirm the importance of hashtags in the Twitter ecosystem, and their prevalence in a variety of contexts. 

\begin{table}[ht]
\caption{{\bf Ranking of popular hashtags.} The first 40 most used hashtags are listed with the corresponding popularity $p$. The hashtags related to the debate and the presidential election such as ledebat, hollande, sarkozy, votehollande, france2012, and présidentielle are recognized.} 
\centering 
\begin{tabular}{c c c c c c} 
\hline\hline 
rank & hashtag & popularity $p$ & rank & hashtag & popularity $p$ \\ [0.5ex] 
\hline 
1 & ledebat & 180946 & 21 & ns & 18715 \\ 
2 & hollande & 143636 & 22 & ps & 18492  \\
3 & sarkozy & 116906 & 23 & teamfollowback & 18476 \\
4 & votehollande &  99908 & 24 & ggi & 17734  \\
5 & radiolondres &  97622 & 25 & bastille & 16056  \\ 
6 & bahrain & 71571 & 26 & présidentielle & 13799 \\
7 & fh2012 & 67759 & 27 & afp & 13710 \\
8 & avecsarkozy & 67549 & 28 & france2 & 12906 \\
9 &  ledébat & 66668 &  29 & syria & 11594 \\
10  &  ff & 49499 &  30 & psg & 10566 \\
11 &  ns2012 & 40337 & 31 & sarko & 10503 \\
12 &  ump & 25125 & 32 & tf1 & 10201\\
13 &  thevoice & 24696 & 33 & mutualite & 10093 \\
14 & fr & 24249 & 34 & egypt & 9970 \\
15 & bayrou & 23029 & 35 & lavictoire & 9949 \\
16 & fh & 22369 & 36 & fn & 9763 \\
17 & rt & 21598 & 37 & franceforte & 9626 \\
18 & france2012 & 20635 & 38 & placeaupeuple & 9211 \\
19 & reseaufdg & 19488 & 39 & jemesouviens & 9098 \\
20 & france & 19268 & 40 & bfmtv & 9010 \\ [1ex] 
\hline 
\end{tabular}
\label{table:nonlin} 
\end{table}

Any type of human activity is influenced by circadian and weekly cycles. This observation has been verified in recent years in a variety of social data sets, going from mobile phone ~\cite{Circadianpatternburstiness_mobilephone} to online social media~\cite{burstydynamicsTwitter,2014arXiv1411.0722F, 2015arXiv150203224M}. In addition, deviations from these cycles can help at detecting atypical events such as responses to catastrophes \cite{scientificrumor, earthquakeTwittercollectiveattention, determinecrisisviahash}. Fig. 1 in the introduction shows the total number of tweets per minute over a sub-period of 6 days and confirms these findings, with clear circadian patterns and two peaks during major public events related to the French presidential election-2012. Besides this smooth periodic behavior, the data also exhibits a noisy signal at a finer time scale, as shown in the inset of Fig. 1.  In the following, we will analyze the properties of this complex time series, by decomposing it into groups of hashtags depending on their popularity, and uncover temporal statistical differences between these groups.

\subsection*{Heterogeneity in popularity of hashtags}

The success of a hashtag can be measured by its popularity $p$, defined as its number of occurrences, and equivalent to its frequency. Fig. 2 presents the Zipf-plot and the probability density function (PDF) of $p$, for the 295,697 unique hashtags observed in the data set. The Zipf-plot [Fig. 2(a)] indicates that more than half of the hashtags ($\approx60\%$) appear just once in the data set, with $p$ = 1. Moreover, around $83\%$ of the hashtags have $p<5$, in the pink-colored region in the last (right) rectangle of Fig. 2(a). For moderate values of $p$, if we set a threshold of $p$ to 1000 with an upper-bound to 25000, only $0.15\%$ of the hashtags fit in the yellow-colored rectangle. Finally, top hashtags with $p>25000$, in the red-colored rectangle, are very rare ($\approx0.0001\%$), but more frequent than would be expected for values so large as compared to the median. These observations are confirmed in Fig. 2(b), where we show the probability distribution of $p$, $P(p)$ in a log-log plot. $P(p)$ is a clear example of a fat-tailed distribution associated with a strong heterogeneity in the system.

\begin{figure}[]
\includegraphics[width= 15.5cm]{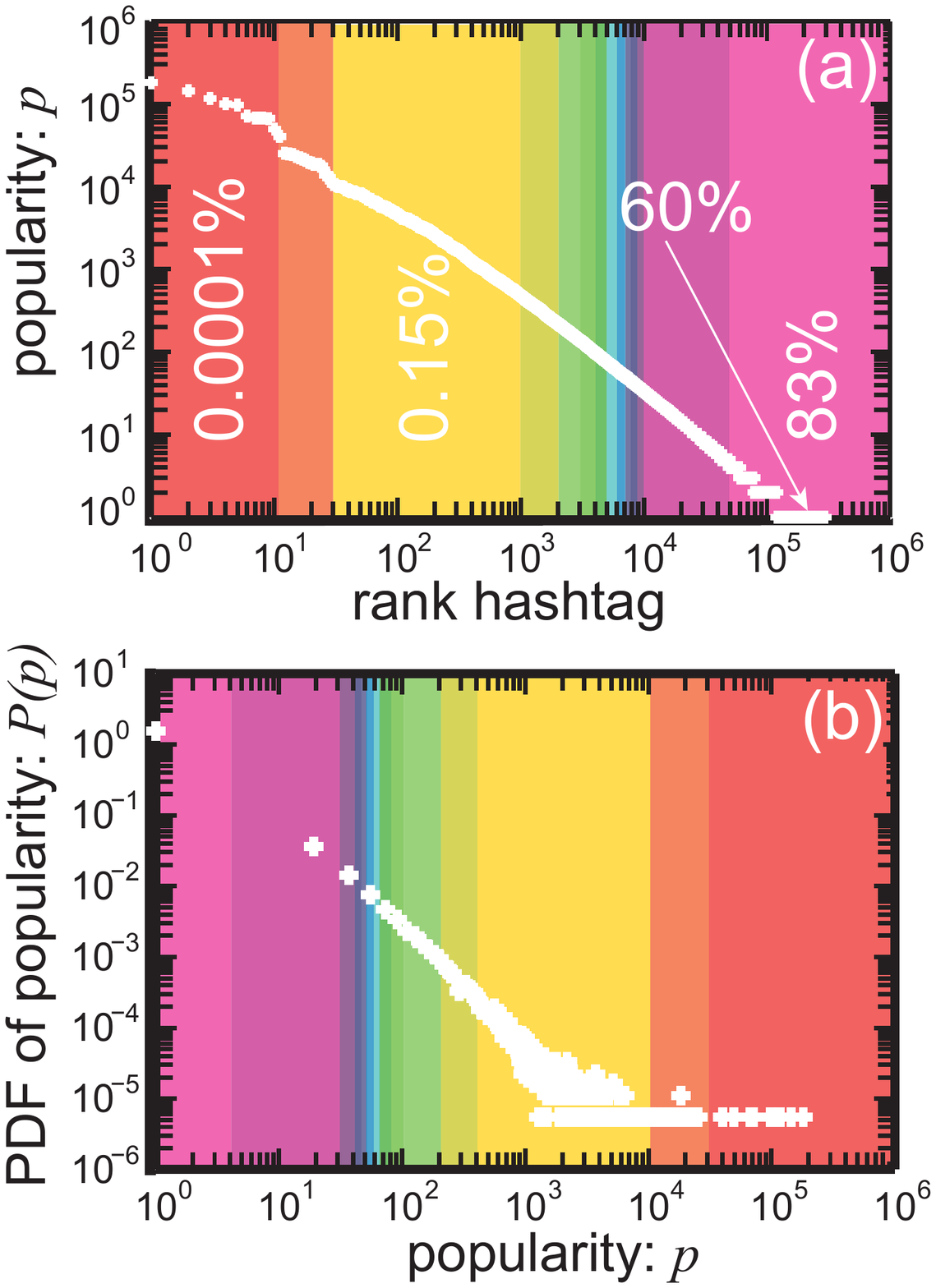}
\caption{Heterogeneity in the hashtag popularity $p$ is shown in (a) Zipf-plot and (b) probability density function (PDF), $P(p)$. (a) Diversity in $p$ (frequency) is visible in a power-law scaling in the log-log plot. We rank hashtag from high $p$ (left) to low $p$ (right). Different colored shaded rectangles highlight the value of $p$ from red and orange (high $p$) to purple and pink (low $p$). The percentages describe the overall contributions of the corresponding rectangles. (b) Similarly, $P(p)$ obeys a slowly decaying function and  presents a power-law distribution with a fat tail. The same colored schema in (a) is applied to visualize the contributions of different values of $p$.} 
\end{figure}

The heterogeneity in $p$ has been already observed~\cite{plane, Viralityprediction_memes, Competition_memes, onlinepopularityheterogeneity}. A mechanism proposed for its emergence is the competition between information overload and the limited capacity of each user~\cite{Competition_memes_limitedattention, Competition_inducedcriticality, Competition_advertisement, 2015arXiv150105956G}, sometimes coupled with cooperative effects~\cite{Competition_memes, cooperationandcompetition}. It has been also shown that hashtags having unique textual features become more popular than hashtags presenting common textual features~\cite{AverageBoring_MemeSuccess}. In this paper, we are not interested in the origin of the heterogeneity, but in its relation with temporal characteristics of hashtags.

\section*{Hashtag spike trains}

\subsection*{Temporal heterogeneity}
We will draw an analogy between hashtag dynamics and neuron spike trains. To this end, we introduce standard methods from spike train analysis into the field of hashtag dynamics. Hashtags are keywords associated to different topics, which can be created, tracked and reused by users. Their popularity and unambiguity make them an essential mechanism for information diffusion in Twitter. The statistical description of neuron spike sequences is essential for extracting underlying information about the brain~\cite{train1}. It was originally believed that in vivo cortical neurons behave as time-dependent Poisson random spike generators, where successive inter-spike intervals are independently chosen from an exponential distribution with a time-dependent firing rate~\cite{train2}. However, more recent observations have shown that the inter-spike interval distribution exhibits significant deviations from the exponential distribution, which has led to the construction of appropriate tools to describe neuron signals~\cite{Lv1, Lv2, Lv3, Lv4, Lv5}. 

Similarly, a hashtag spike train is defined as the sequence of timings at which a hashtag is observed in Twitter. In this framework,  we do not specify the type of dynamics of hashtags, endogeneous or exogeneous~\cite{Dynamicsofbooksale}, i.e. endogeneous, hashtag diffusion among members of the social network, or exogeneous, the diffusion driven by external factors such as TV and newspapers, but only in the timings. Each hashtag thus generates a unique hashtag spike train with a characteristic popularity $p$. As a first basic indicator, in Figs. 3(a,b) we show the inter-hashtag spike interval cumulative and probability distributions, $CDF(\Delta\tau)$ and $P(\Delta\tau)$, respectively. In order to avoid artificially deforming the distributions because of heterogeneity in $p$, we classify $CDF(\Delta\tau)$ and $P(\Delta\tau)$ in classes depending on $p$, illustrated by different colors in Fig. 2. We observe similar behavior across the classes, as
$P(\Delta\tau)$ deviates strongly from an exponential distribution (Poisson), $P(\Delta\tau)$ = $\xi e^{-\xi\Delta\tau}$, where $\xi$ is a firing rate (frequency and so $p$ in our concept) at which hashtags appear. Instead, we observe fat-tailed distributions~\cite{scientificrumor, onlinepopularityheterogeneity, BarbarasiOriginofBursts, TaroNaokiVotermodel_IEI, Circadianpatternburstiness_mobilephone, ABarratHeterogeneityIEI, Altmann_extreme_events} as shown in Fig. 3(b) for high and moderate $p$. As mentioned in the introduction, this deviation may either originate from temporal correlations or non-stationary patterns, making the system different from a stationary, uncorrelated random signal.

\begin{figure}[]
\includegraphics[width= 8cm]{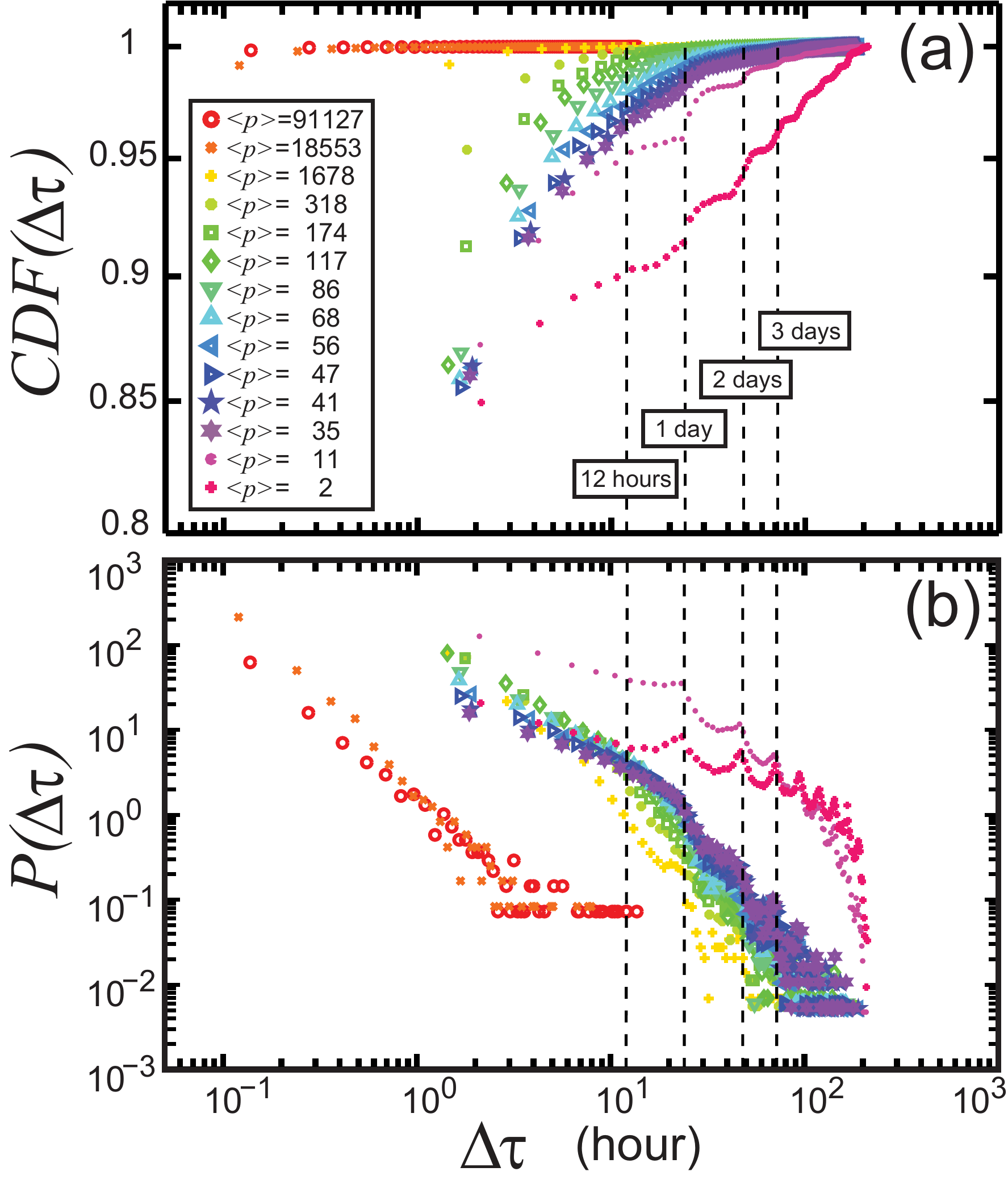}
\caption{The cumulative (a), $CDF(\Delta\tau)$, and probability (b), $P(\Delta\tau)$, distributions of the inter-hashtag spike intervals. We observe that $P(\Delta\tau)$ exhibits, for different classes of hashtags distinguished by their popularity, non-exponential features. The different colors correspond to those in Fig. 2. The legend provides the average popularity $\langle p\rangle$ in each hashtag class. The dash lines indicate the positions of 1 day, 2 days, and 3 days, where $P(\Delta\tau)$ gives peaks for low $p$ (pink symbols). The binning is varied from 8 minutes to 2 hours depending on $p$, e.g. 8 min. for high $p$ (red-orange), 1.5 hour for moderate $p$ (yellow-green-blue-purple), and 2 hours for low $p$ (pink). All $P(\Delta\tau)$ present maxima at 1 second, which is not shown to describe tails in a larger window.}  
\end{figure}

\subsection*{Real and randomized data sets}

We will analyze two sets of data, which we now describe: The empirical data set, directly coming from the data, and a randomized data set, serving as a null model in our analysis. 

The \textit{real data set} contains one spike train per hashtag, as illustrated in Fig. 4(a). The time resolution of the spikes is the same as that of the data set, that is 1 second. In situations when multiple spikes of the same hashtag take place at the same time only one event is considered. The statistics of such events are provided at the end of this subsection. In each spike train, the appearance time of the spikes is ordered from the earliest time to the latest time. 

\begin{figure}[]
\includegraphics[width= 16.5cm]{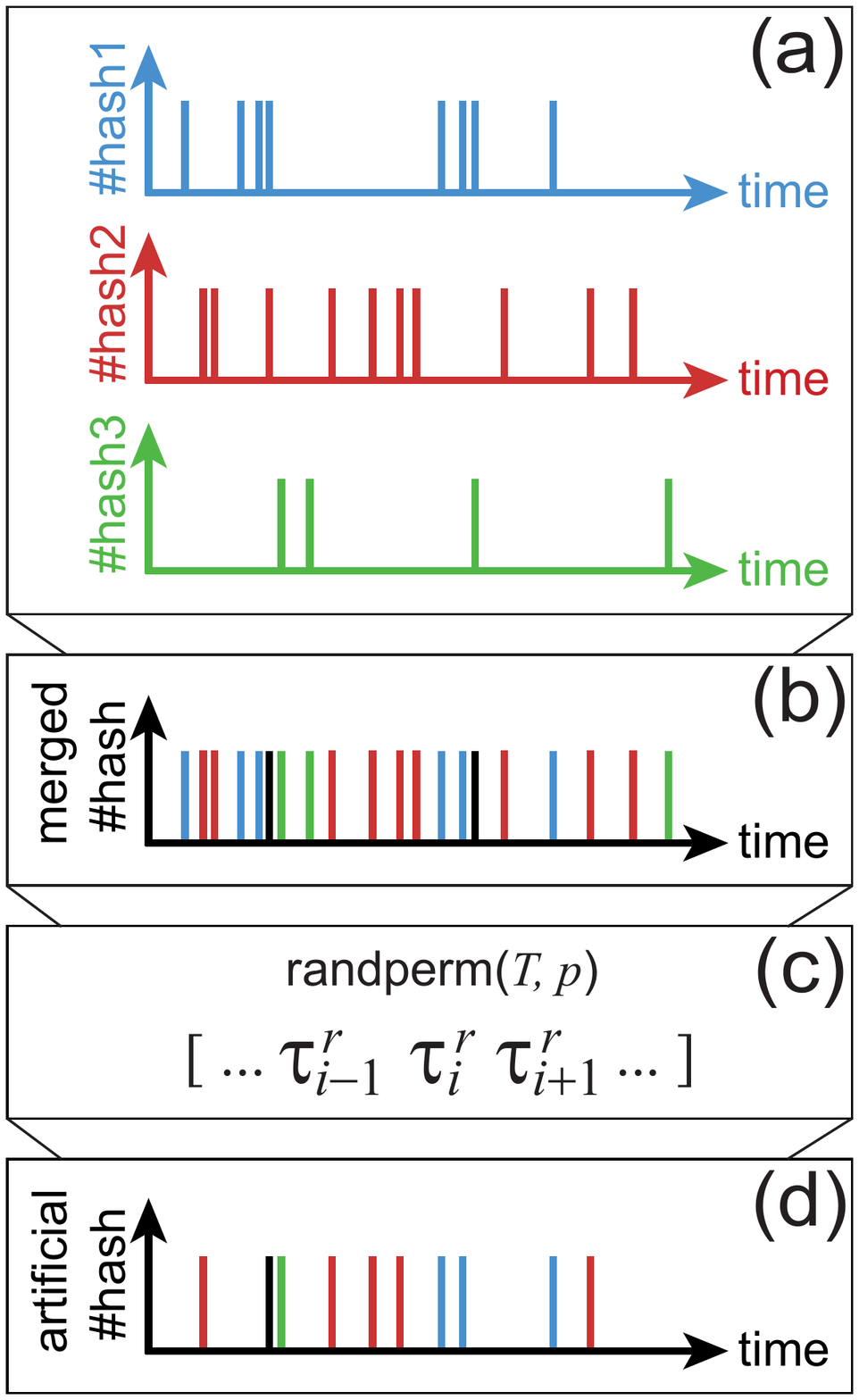}
\caption{Real and artificial hashtag spike trains. (a) As an illustration of different hashtag spike trains representing different types of hashtag propagation of the data set. (b) Merging hashtag spike trains from the real data. The black spikes describe that only one activity is counted if multiple activities occur at the same time. (c) Randomization procedure by randperm (Matlab). $T$ contains full hashtag activity of the data set. The randperm gives a matrix $p$, unique independent numbers out of $T$, and constructing random time series $\ldots$, $\tau^r_{i-1}$, $\tau^r_i$, $\tau^r_{i+1}$, $\ldots$ from full hashtag activity matrix $T$. (d) The resultant artificial hashtag spike train.}
\end{figure}

The \textit{random data set} is randomized version of the real data set, where each spike train of size $p$ generates a spike train of the same size with random times. In practice, we first combine all hashtag spike trains and obtain one merged hashtag spike train as illustrated in Fig. 4(b). This train carries the full history of all hashtags and, importantly, reproduces the nonstationary features of the original data in the presence of temporal correlations, burstiness, and the cyclic rhythm. As before, if two or more spikes generated in the same time, only one spike is shown in that time in the merged spike train, e.g. see the black spikes in Fig. 4(b). 

Randomization is performed by permuting elements, as shown in Fig. 4(c), for instance by using randperm($T$, $p$) in Matlab. Here, $T$ represents the full matrix of times in the merged spike train and $p$ is the desired popularity, number of total spikes in a train. The permutation procedure generates $p$ times uniformly distributed unique numbers out of $T$ and these numbers define the artificial spike train, e.g. $\ldots$, $\tau^r_{i-1}$, $\tau^r_i$, $\tau^r_{i+1}$, $\ldots$, as shown in Fig. 4(c). In our data set, $p\ll T$ is always verified, as the maximum $p$ is 180,900 and the length of $T$ is 667,996. This procedure is applied to each spike train of size $p$ [Fig. 4(d)]. Generating independent, yet time-dependent events, the procedure is expected to create time-dependent Poisson random processes, $P(\Delta\tau,t)$ = $\xi(t)e^{-\xi(t)\Delta\tau}$, where the firing rate $\xi(t)$ in this case explicitly depends on the time of the day and of the week.

\textit{Statistics of multiple tweets in 1 second.} We detect multiple occurrences in 1 second for 6661 hashtags. Fig. 5 presents the probability distribution $P(c_h)$ of observing $c_h$, occurrences of an hashtag during one second, for different hashtag popularity class. Even though $c_h>1$ occurs rarely, we observe that this possibility is more probable for popular hashtags (red open circles), as expected. For the most popular hashtag, ledebat, one finds $max(c_h)$ = 40.

\begin{figure}[]
\includegraphics[width= 8cm]{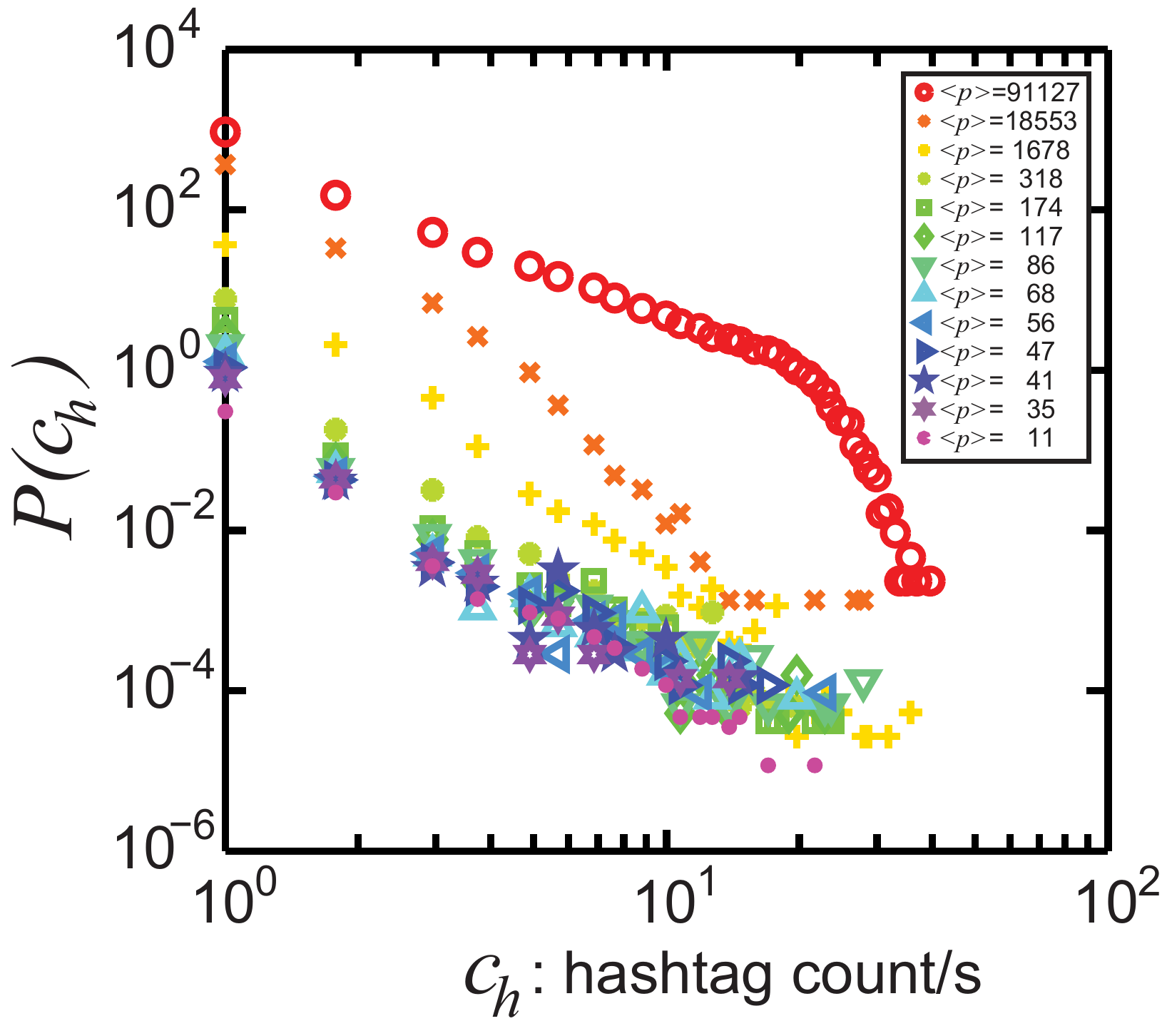}
\caption{The probability distribution of count of hashtag activity per second $P(c_h)$. We show that, except for the top most popular hashtags listed in Table 1 with ranking 1-11 and presented here in red symbols, multiple activity in 1 second is very rare. The different colors correspond to those in Figs. 2 and 3. The legend provides the average popularity $\langle p\rangle$ in each hashtag class.}
\end{figure}

\section*{Local variation}

The time series of spike trains are inherently nonstationary, as shown in Fig. 1. For this reason, metrics defined for stationary processes are inadequate and might lead to incorrect conclusions. For instance, the non-exponential shapes of the inter-event time distribution $P(\Delta\tau)$ in Fig. 3 might originate either from correlated and collective dynamics, or from the nonstationarity of the hashtag propagation. Similarly, statistical indicators based on this distribution, such as its variance or Fano factor, might be affected in a similar way. For this reason, we consider here the so-called local variation $L_V$, originally defined to determine intrinsic temporal dynamics of neuron spike trains~\cite{Lv1, Lv2, Lv3, Lv4, Lv5}. 

Unlike quantities such as $P(\Delta\tau)$, $L_V$ compares temporal variations with their local rates and is specifically defined for nonstationary processes~\cite{Lv5}
\begin{equation}
L_V=\frac{3}{N-2}\sum\limits_{i=2}^{N-1} \left(\frac{(\tau_{i+1}-\tau_i)-(\tau_{i}-\tau_{i-1})}{(\tau_{i+1}-\tau_i)+(\tau_{i}-\tau_{i-1})}\right)^2 
\label{Eq:Lv_tau}
\end{equation}
Here, $N$ is the total number of spikes and $\ldots$, $\tau_{i-1}$, $\tau_i$, $\tau_{i+1}$, $\ldots$ represents successive time sequence of a single hashtag spike train. Eq.~\ref{Eq:Lv_tau} also takes the form~\cite{Lv5}
\begin{equation}
L_V=\frac{3}{N-2}\sum\limits_{i=2}^{N-1} \left(\frac{\Delta\tau_{i+1}-\Delta\tau_{i}}{\Delta\tau_{i+1}+\Delta\tau_{i}}\right)^2
\label{Eq:Lv_Deltatau}
\end{equation}
where $\Delta\tau_{i+1}$ = $\tau_{i+1}-\tau_i$ and $\Delta\tau_{i}$ = $\tau_{i}-\tau_{i-1}$.  $\Delta\tau_{i+1}$ quantifies forward delay and $\Delta\tau_{i}$ represents backward waiting time for an event at $\tau_{i}$. Importantly, the denominator normalizes the quantity such as to account for local variations of the rate at which events take place. By definition, $L_V$ takes values in the interval [0:3]. 

The local variation $L_V$ presents properties making it an interesting candidate for the analysis of hashtag spike trains~\cite{Lv1, Lv2, Lv3, Lv4, Lv5}. In particular, $L_V$ is on average equal to 1 when the random process is either a stationary or a non-stationary Poisson process~\cite{Lv1}, with the only condition that the time scale over which the firing rate $\xi(t)$ fluctuates is  slower than the typical time between spikes.
Deviations from 1 originate from local correlations in the underlying signal, either under the form of pairwise correlations between successive inter-event time intervals, e.g. $\Delta\tau_{i+1}$ and $\Delta\tau_{i}$ which tend to decrease $L_V$, or because the inter-event time distribution is non-exponential. An interesting case is given by Gamma processes~\cite{Lv1, Lv3}
\begin{equation}
P(\Delta\tau, t; \xi, \kappa) = (\xi \kappa)^\kappa\Delta\tau^{(\kappa-1)} e^{-\xi \kappa\Delta\tau}/\Gamma(\kappa)
\end{equation}
where 
$\kappa$ is called a shape parameter and determines the shape of the distribution and $\Gamma$ is the Gamma function. Here, $\xi$ and $\kappa$ are the two parameters of the Gamma process. While $\xi$ determines the speed of the dynamics, $\kappa$ controls for the burstiness (irregularity) of the spike trains. Assuming that events are independently drawn, the shape factor is related to $L_V$ as follows~\cite{Lv1, Lv3}
\begin{equation}
\langle L_V\rangle = \frac{3}{2 \kappa +1}
\end{equation}
Here, the brackets describe the average taken over the given distribution~\cite{Lv1}. 
When $\kappa$ = 1, an exponential is recovered, and one finds $\langle L_V\rangle$ = 1 as expected. Smaller values of $\kappa$ increase the variance in $\Delta\tau$ and therefore its burstiness, making $L_V$ larger than 1. On the other hand, larger values of  $\kappa$ decrease the variance of $\Delta\tau$ and the burstiness of the process, making $\langle L_V\rangle\approx0$ smaller than 1.

We measure $L_V$ of hashtag spike trains and group the values depending on the popularity $p$ of their hashtag as was done in Figs. 2 and 3. Fig. 6 shows scatter plots of $L_V$ for the real data set (a), the empirical sequence $\ldots$, $\tau_{i-1}$, $\tau_i$, $\tau_{i+1}$, $\ldots$, and the random data set (b), the random sequence $\ldots$, $\tau^r_{i-1}$, $\tau^r_i$, $\tau^r_{i+1}$, $\ldots$, on linear-log plots. Different colors are used to distinguish the different groups and the inset legend provides the average popularity $\langle p\rangle$ in the groups. 

\begin{figure*}[]
\includegraphics[width= 16cm]{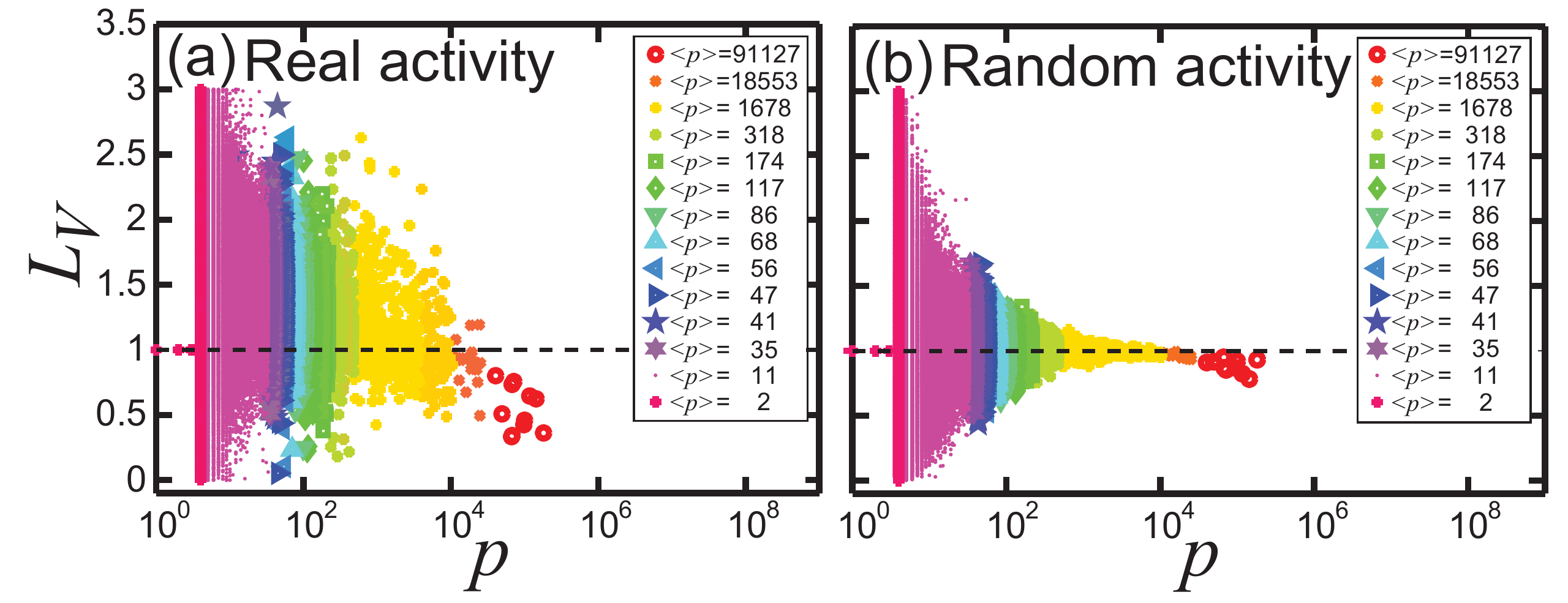}
\caption{The local variation $L_V$ of hashtag spike trains versus popularity $p$ on a linear-log plot. Each color and symbol summarized in the legend present different range of $p$: Low $p$, pink and purple colors, and moderate $p$, blue, green, and yellow colors, and then high $p$, orange and red colors. In addition, the average $p$, $\langle p\rangle$, indicated in the legend ranks colors and symbols quantitatively. (a) Hashtag spike trains of the data set. (b) Artificial hashtag spike trains.}
\end{figure*}

A more readable representation is provided in Fig. 7, where we show histograms $P(L_V)$ of the values of $L_V$, for the two data sets and for the distinguished hashtag groups in $p$. The results clearly show that $L_V$ fluctuates around 1 in the random data set, as expected for a time-dependent Poisson process. On the other hand, $L_V$ systematically deviates from 1 in the original data set, where temporal correlations are present. 

\begin{figure}[]
\includegraphics[width= 8cm]{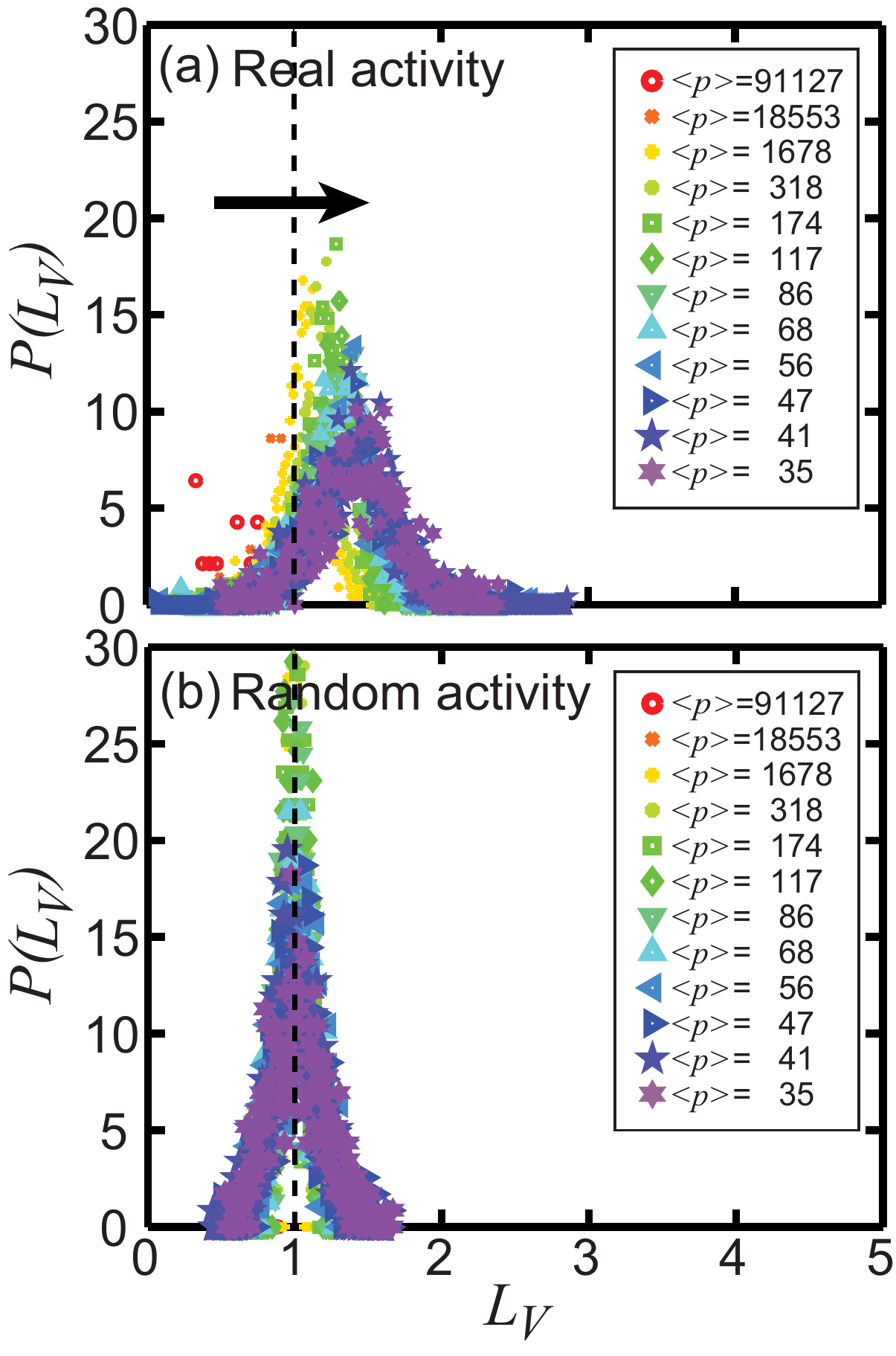}
\caption{Probability density function (PDF) of the local variation $L_V$ of real hashtag propagation (a) and random hashtag time sequence (b). Two distinct shapes are visible: (a) From high $p$ to low $p$, the peak position of $P(L_V)$ shifts from low values of $L_V$ to higher values of $L_V$. (b) $P(L_V)$ always peaks around 1 for the random sequences generated by artificial hashtag spike trains. The same color coding is applied as used in Fig. 6.} 
\end{figure}

The observation is confirmed in Fig. 8(a), where we plot the mean $\mu(L_V)$ of $L_V$, with error bars, as a function of $\langle p\rangle$. Furthermore, $L_V$ of the original data indicates that high impact hashtags (high $p$) are associated with lower values of $L_V$ suggesting more homogeneous (regular) time distributions. These results confirm the potential use of $L_V$ as a metric to capture deviations from Poisson (temporarily uncorrelated) processes, but also to identify distinct statistical properties generated specifically in high $p$. Moreover, Fig. 8(b) presents the statistical differences between the real and the random spike trains in detail. The deviations from Poisson processes where $\mu_{0}(L_V)$ = 1 are calculated by $z$ = $\mu(L_V)-\mu_{0}(L_V)/\sigma(L_V)/\sqrt{n}$ with the standard deviations of $L_V$, $\sigma(L_V)$, and the number of the data points given in the distributions in Fig. 7, $n$. We observe that $z-$values for the random spikes are almost equal to 0, excluding in high $p$, indicating the agreement between Poisson signals and our random spike trains, which is not the case for the real trains giving $z\ncong0$ in any of $\langle p\rangle$. 


To conclude, we perform an analysis to test the persistence of the temporal characteristics of hashtags, as measured by $L_V$, through time. To do so, we divide each hashtag time series into two time series. The resulting values of local variations are $L_V(t_1)$ for the first half of a spike train and $L_V(t_2)$ for the second half of the train, and then we calculate the Pearson correlation coefficient $r(L_V(t_1), L_V(t_2))$ between these values~\cite{LVcorrelation}.  In Fig. 9(a) we show the linear relation between $L_V(t_1)$ and $L_V(t_2)$.  Fig. 9(b) shows $r(L_V(t_1), L_V(t_2))$ as a function of the average popularity $\langle p\rangle$. Both indicate that values of $L_V$ for the same hashtag at different times is significantly and temporarily correlated. We also observe that bursty (low $p$) and regular (high $p$) signals give small $r$, while the spike trains with moderate $p$ provide the largest values of $r$ where $L_V$ suggests more uniform temporal behavior through the individual trains. 

\begin{figure}[]
\includegraphics[width= 9cm]{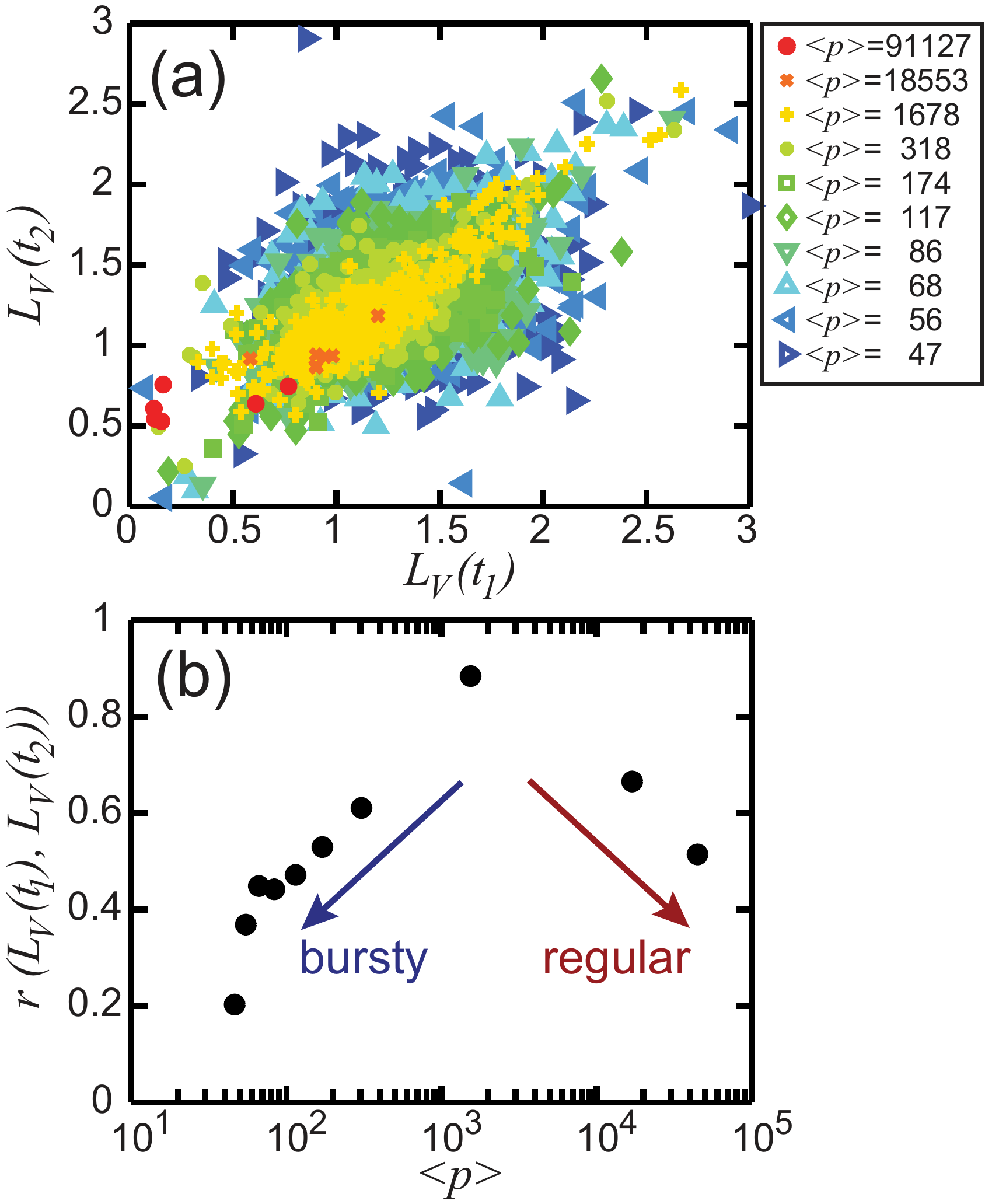}
\caption{Linear correlation of $L_V$ through real hashtag spike trains. (a) The linear relation of the first and the second halves of the empirical spike trains, $L_V(t_1)$ and $L_V(t_1)$, respectively, are investigated. The legend ranks $\langle p\rangle$ in different colors and symbols. (b) The Pearson correlation coefficient $r(L_V(t_1), L_V(t_2))$ between these quantities show that while the temporal correlation through moderately popular hashtag is maximum, $r$ reaches the minimum values for both bursty (high $L_V$ and low $p$) and regular (low $L_V$ and high $p$) spike trains.}
\end{figure}

\section*{Discussion}

The main purpose of this paper is to introduce a statistical measure suitable for the analysis of nonstationary time series, as they often take place in online social media and communications in social systems. As a test case, we have focused on the dynamics of hashtags in Twitter. However, the same methodology could be also applied to the other types of correlated, bursty, and nonstationary signals, for instance the dynamics of cascades in Twitter and Facebook or phone call activity. 

Instead of measuring standard statistical properties of noisy hashtag signals such as the inter-event time distribution, its variance or the Fano factor, conventionally applied to characterize the burstiness of a signal, we have focused on the local variation $L_V$, a metric capturing the fluctuations of the signal as compared to a local characteristic time. This measure, previously defined for neuron spike train analysis, nicely uncovers the regularity and the firing rate of the trains~\cite{Lv1, Lv2, Lv3, Lv4, Lv5} and so helps to identify local temporal correlations. It is important to stress that the current analysis exclusively focuses on properties of time series, and does considers neither the mechanisms leading to the observed statistical dynamic properties nor the effect of the underlying topology, e.g. through following-follower relations. An interesting line of research would study the relation between $L_V$, the underlying topology~\cite{Uncoveringtemporaldynamics} and diffusive models, for instance Hawkes process~\cite{Burstingpointprocess, Hawkespointprocesses}. In addition, both neurons~\cite{train2} and hashtags can be driven by multiple firing rates and $L_V$ analysis associated to Gamma distributions would provide more concrete results on hashtag spike trains, as done for neuron spikes~\cite{Lv3}. 

We should also note that the finite temporal resolution of the data (1 sec), and the fact that multiple events per time window are neglected, tends to make $L_V$ artificially decrease for popular hashtags. In an extreme case, the time series is indeed regular, with events taking place every second. In this work, we have therefore carefully verified that fluctuations in $L_V$ are not artificially driven by these limitations. To do so, we have compared the values of $L_V$ in the empirical data with those of a null model. We observe a small decay of $L_V$ for popular hashtags in the null model (see Fig. 8), but this decay is much more limited than the one observed in the empirical data, e.g. $L_V$ = 0.89 for $\langle p\rangle$ =  $10^5$ in the null model while it is equal to  $L_V$ = 0.54 for the real data. In addition, a decay of $L_V$ in real hashtag data is also present in moderately popular hashtags, where multiple events per second are very rare.  An interesting research direction would be to generalize the definition of local variation in order to allow for the analysis of multiple events per time window, thereby evaluating the deviations of dense time series to non-stationary Poisson processes. Finally, in a finite time window, as observed in empirical data, the statistics of high frequency hashtags is much better than that of low frequency hashtags, simply because the former occur many more times than the latter. For this reason, measurements of $L_V$ for low popularity hashtags are more subject to noise.

The empirical analysis also reveals an interesting pattern observed in the data, as more popular hashtags tend to present a more regular temporal behavior. This lack of burstiness ensures that hashtags do not disappear from the social network for very long periods of time, thereby allowing for a regular activation of the interest of Twitter users. These findings are reminiscent of recent observation in numerical simulations showing that
burstiness hinders the size of cascades~\cite{PhysRevE.89.062815}, and 
should be incorporated into the modeling of theoretical information diffusion models, in particular threshold~\cite{Karimi20133476} and stochastic~\cite{Kawamoto20133470} models, on temporal networks but also into the ranking models capturing online heterogeneity in the empirical data~\cite{onlinepopularityheterogeneity}. 

\section*{Supporting Information}

{\bf Supporting Information S1 Supporting data files.}\\
\noindent (Compressed zip folder of .dat files - will be publicly available online)

\begin{acknowledgments}
We thank Takaaki Aoki and Taro Takaguchi for their useful comments and Lionel Tabourier for providing data set. C. Sanl{\i} acknowledges supports from the European Union 7th Framework OptimizR Project, FNRS (le Fonds de la Recherche Scientifique, Wallonie, Belgium), and National Institute of Informatics, Tokyo, Japan. This paper presents research results of the Belgian Network DYSCO (Dynamical Systems, Control, and Optimization), funded by the Interuniversity Attraction Poles Programme, initiated by the Belgian State, Science Policy Office.
\end{acknowledgments}

\bibliography{refs_Lvhashtagspiketrains_FULLinfo}

\end{document}